\documentclass{article}
\usepackage{spconf,amsmath,graphicx}
\usepackage{graphicx}
\usepackage{amsfonts}
\usepackage{multirow}
\usepackage{float}
\usepackage{array, tabularx}
\usepackage{booktabs}
\usepackage{lipsum}
\usepackage{subcaption}
\usepackage{dblfloatfix}

\usepackage{color}

\title{Non-intrusive Speech Quality Assessment Using Neural Networks}
\name{\em Anderson R. Avila$^*$$^1$, Hannes Gamper$^2$, Chandan Reddy$^3$, Ross Cutler$^3$, Ivan Tashev$^2$, Johannes Gehrke$^3$ \thanks{$^*$Work on this project performed as an intern at Microsoft Research Labs, Redmond, WA.}}

\address{$^1$Institut National de la Recherche Scientifique, Montreal, QC, Canada\\
  $^2$Microsoft Research Labs, Redmond, WA, USA\\
  $^3$Microsoft Corporation, Redmond, WA, USA\\
 \emph{anderson.avila@emt.inrs.ca, \{hagamper, chkarada, rcutler, ivantash, johannes\}@microsoft.com}}

\begin{document}
%
\maketitle
%


\begin{abstract}

\noindent Estimating the perceived quality of an audio signal is critical for many multimedia and audio processing systems. Providers strive to offer optimal and reliable services in order to increase the user quality of experience (QoE). In this work, we present an investigation of the applicability of neural networks for non-intrusive audio quality assessment. We propose three neural network-based approaches for mean opinion score (MOS) estimation. We compare our results to three instrumental measures: the perceptual evaluation of speech quality (PESQ), the ITU-T Recommendation P.563, and the speech-to-reverberation energy ratio. Our evaluation uses a speech dataset contaminated with convolutive and additive noise, labeled using a crowd-based QoE evaluation, evaluated with Pearson correlation with MOS labels, and mean-squared-error of the estimated MOS. Our proposed approaches outperform the aforementioned instrumental measures, with a fully connected deep neural network using Mel-frequency features providing the best correlation ($0.87$) and the lowest mean squared error ($0.15$).


\end{abstract}
\begin{keywords}
Audio quality assessment, speech quality assessment, deep neural network
\end{keywords}
\section{Introduction}
\label{sec:intro}

\begin{figure*}
    \centering
    \includegraphics{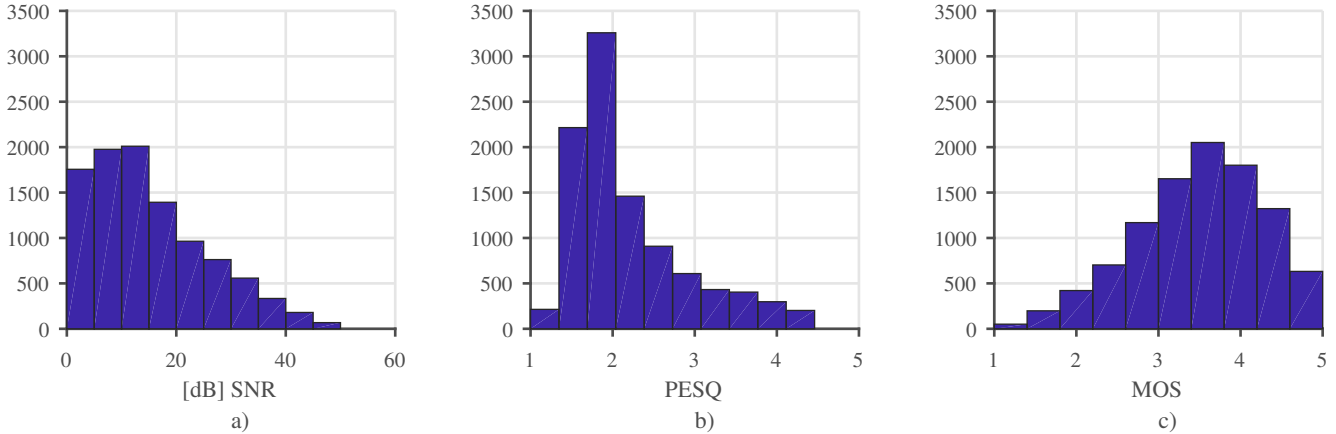}
    \caption{Histogram for the distribution of (a) SNR, (b) PESQ, and (c) MOS.}
    \label{fig:distribution}
\end{figure*}

In speech communication systems, the audio signal can be affected by background noise, reverberation, enhancement algorithms as well as by network impairments. In such scenarios, as providers strive to guarantee optimal and reliable services to their customers, estimating the perceived quality of the audio signal has become crucial. For instance, speech quality prediction can be useful during network design and development as well as for monitoring and improving customers' quality of experience (QoE) \cite{falk2006}.

The subjective listening test is the most accurate method for evaluating perceived speech signal quality \cite{ITU_T_P_800}. In this approach, the estimated quality is the average of users' judgment, usually in a scale ranging from 1 to 5. The average of all participants' scores over a specific condition is referred to as the mean opinion score (MOS) and represents the perceived speech quality after leveling out individual factors \cite{falk2011}. Such subjective measurements are not always feasible as they: (1) require a considerable number of listeners; (2) can be laborious and time-consuming; (3) can be expensive; and (4) perhaps, more importantly, cannot be done in real-time \cite{avila2016}.

As an alternative, several objective instrumental quality measures have been proposed and standardized. The ITU-T Recommendation P.862, referred to as Perceptual Evaluation of Speech Quality (PESQ) \cite{ITU_T_P862}, is one of the most widely used measures for audio quality assessment, followed by its improved version ITU-T Recommendation P.863, also known as Perceptual Objective Listening Quality Assessment (POLQA) \cite{ITU_T_P863}. These models, however, were developed specifically for distortions introduced by speech compression (i.e., codecs) and show low performance when the audio signal is corrupted by noise, reverberation, or processed by new enhancement algorithms \cite{falk2015}.

In addition, many of these approaches are intrusive as they require the reference clean speech signal to estimate the MOS. This limits their application to use with a synthetic dataset. There are numerous algorithms and standards which are non-intrusive \cite{GranchovZhaoLindblomEtAl2006}; they use only the corrupted speech signal for quality assessment. Normally, these algorithms are expected to be less accurate in estimating the perceptual sound quality.

Despite the breakthroughs of neural networks in so many areas, to date, only a handful of neural network-based models have been proposed \cite{soni2016novel,spille2018predicting, fu2018quality}. To the best of our knowledge, even the most recent methods to predict MOS present serious limitations. First, most of them are developed to measure intelligibility \cite{spille2018predicting}, which is just one aspect of audio quality \cite{falk2011}. Second, these neural network-based models are trained on a limited number of conditions, usually with no interaction between different impairments, which is quite unrealistic and rarely happens in everyday scenarios. Finally, the ground truth to train these models frequently are not the subjective scores (MOS), but the scores of another model, such as PESQ \cite{fu2018quality}, which leaves out most of the relevant human factors.

To tackle these limitations, we generated a realistic dataset, and we labeled it using a crowd-based QoE estimation \cite{hossfeld2014best}. We explore three neural network-based architectures to predict MOS. In the first approach, we use a psychoacoustic inspired feature, namely the constant Q transform \cite{todisco2017constant, brown1991calculation}, which has been successfully adopted to distinguish natural and unnatural speech as well as to perform music analysis. These features are used as input to a convolutional neural network (CNN). In the second approach, we explore the low-dimensional total variability (TV) space \cite{dehak2011front}. The features projected in the TV space, also referred to as i-vectors, are then used as input to a fully connected deep neural network (DNN). The third approach is based on the Mel-frequency features, combined with a DNN. The performance of the proposed approaches are compared with three instrumental measures: PESQ, the ITU-T Recommendation P.563, and the speech-to-reverberation energy ratio (SRMR).

The remainder of this document is organized as follows. In Section $2$, we present the data generation process. The features adopted in this work are described in Section $3$. Section $4$ presents the neural network models evaluated in this paper. Our experimental setup is described in Section $5$, and results are discussed in Section $6$. Section $7$ concludes the paper.

\section{The Audio Quality Evaluation Dataset}
\label{sec:quality}

 \begin{figure*}
   \centering
    \includegraphics[width=0.9\textwidth]{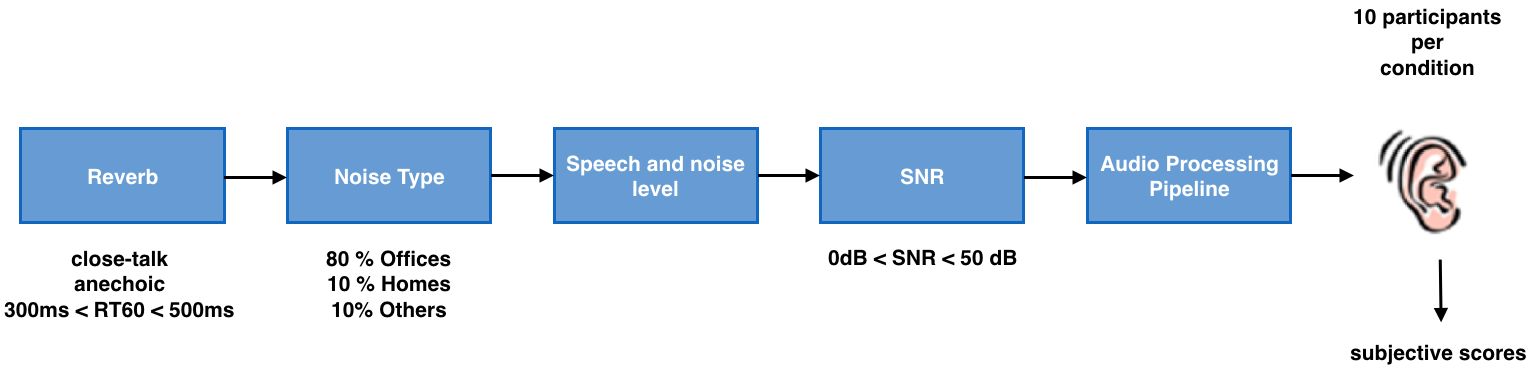}
    \caption{Block diagram describing the distortions introduced in the audio signal and labeling by 10 participants.}
    \label{fig:quality}
\end{figure*}



In everyday environments, it is expected that an audio signal is subject to a variety of acoustic background distortions. To create realistic scenarios for our listening quality test, we created a dataset of 10,000 samples, representing the conditions to be assessed. We first generated $2,010$ clean speech files, equally distributed by gender: $670$ males, $670$ females and $670$ children. Each speech file is approximately $20$ seconds long, starting with $4$ seconds of silence, followed by $3$ utterances, separated by $2$ seconds of silence. All samples are normalized to $-23$~dB~FS, and are sampled with $16$~kHz. The human voice levels were modeled with a mean of $65$~dB~SPL at $1$~meter, and deviation of $8$~dB. Then the clean signal is convolved with a randomly selected room impulse response (RIR) from a library of 120~RIRs. They were measured at distance between the source (speakers) and target (microphones) varying between $0.5$ and $3$~meters, in rooms with $RT_{60}$ ranging from $300$ to $500$~ms. Anechoic and close-talk microphone conditions are also included. Next, noise is added to the convolved audio signal, with a mean level of $45$~dB~SPL and deviation of $15$~dB. Three types of noise were considered: offices ($80$\%), homes ($10$\%) and others ($10$\%). The ratio of office noise is higher as it is a more prominent noise type in our use case. The resulting SNRs were limited to $[0, 50]$~dB, as depicted in Fig.~\ref{fig:distribution}a. Finally, half of the corrupted samples were processed with an audio processing pipeline, consisting of a noise suppressor and automatic gain control (AGC). This allowed us to investigate how processed and unprocessed speech are perceived by human users.


As listening room and equipment may influence the outcome of the experiment, listening quality tests are commonly performed with each participant using the same listening conditions, usually a quiet chamber of controlled dimensions \cite{ITU_R_BS}. This, however, does not represent realistic scenarios encountered in real life. Thus, we used crowd-sourcing to label our data. In this type of experiment, online workers are assigned to the task, which can be undertaken in a variety of ambiance \cite{hossfeld2014best} and listening devices. Before initiating the experiment, participants were submitted to a training phase where they listened at least once, but if necessary as many times as they wanted, to samples of each impairment. This was meant to familiarize the participants to the most uncommon distortions and evaluation scale.  After the training, the labelers had to pass a mandatory qualification step. They were asked to rate gold-standard samples. Only participants who successfully passed the qualification were considered for the experiment. Fig. \ref{fig:quality} summarizes the dataset generation and labeling procedure. The perceptual audio quality of each audio sample was rated by ten judges, and the MOS is computed by averaging the scores. Fig.~\ref{fig:distribution}-b and -c provide the histograms of the computed PESQ and averaged MOS. It is well visible that PESQ gives lower scores than human judges.

\section{Proposed Features}
\label{sec:features}
This section describes the three features adopted as input to our neural network models. We first present the constant Q spectrum, then the low-dimensional total variability space, and finalize with Mel-frequency features.

The short-time Fourier transform (STFT) is the most popular time-frequency representation of an audio signal. To extract it, one must choose a short window function that will be multiplied along the audio signal. The length of the window function is fixed, and commonly set to values between $10$ and $30$~ms. The quality factor $Q_c$ \cite{todisco2017constant} for the center of the frequency band $f_c$ is defined as:

\begin{equation}
Q_c = \frac{f_c}{\delta_f},
\end{equation}

\noindent where $\delta_f$ is the frequency bandwidth. Note that for fixed width the quality factor increases with increasing center frequency. This is not aligned well with human perception, which is known to have a constant Q factor between $500Hz$ and $20kHz$ \cite{todisco2017constant}. Perceptually motivated, the constant Q transform (CQT) was introduced in \cite{youngberg1978constant} and later refined in \cite{kashima1985bounded}. Applying the CQT allows better time-frequency resolution as described in \cite{todisco2017constant}. Inspired by this, we include the constant Q spectral in the set of features to be evaluated. The feature dimension is 240x220. In the case of short duration utterances, the last frame was replicated the number of times necessary to attain 220 frames. Also, exceeding frames were removed from long utterances. This procedure was performed to assure that the CNN had always the same input size. This configuration was chosen empirically based on the average duration of the speech files.


The i-vector framework maps a list of feature vectors, $O=\{o_t\}_{t=1}^N,$ where $o_t\in\mathbb{R}^F$, and $N$ is the frame index. Typically Mel-frequency cepstral coefficients (MFCC's) extracted from an utterance, into a fixed-length vector, $n\in\mathbb{R}^D$. In order to achieve that, a Gaussian mixture model (GMM), $\lambda=(\{w_k\},\{m_k\},\{\sigma_k\})$, is used. The GMM, trained on multiple utterances, is referred to as the universal background model (UBM), and is used to collect Baum-Welch statistics from each utterance \cite{garcia2011analysis}. Such statistics are computed for each mixture component $k$, resulting in the so-called supervector $M\in\mathbb{R}^{FK}$, where $F$ represents the feature dimension and $K$ is the number of Gaussian components. As in the Joint Factor Analysis (JFA) \cite{kenny2007joint}, the i-vector framework also considers that speaker and channel variability lies in a lower subspace of the GMM supervectors \cite{hansen2015speaker}. The main difference between the two approaches is that the i-vector projects both speaker and channel variability into the same subspace, namely total variability space, represented as follows:

\begin{equation}
    M = m + Tw,
\end{equation}

\noindent where $M$ is the dependent supervector (extracted from a specific utterance) and $m$ is the independent supervector (extracted from the UBM), $T$ corresponds to a rectangular low-rank total variability matrix and $w$ is a random vector with a normal distribution, the so-called i-vector. In our experiments, a 400-dimensional i-vector was adopted.


To extract Mel features, the audio signal is processed in frames of $512$~samples, with a step size of $160$~samples, at a sampling rate of $16$~kHz. For each frame, $26$~Mel-frequency cepstral coefficients (MFCCs) are computed. The MFCCs are combined with pitch estimate, the output of a voice activity detector (VAD), and the log-power energy of the frame, as well as their first derivatives estimated using the preceding frame. The input to the neural network consists of the features computed for each speech-active frame, as determined by the VAD, plus the 12 frames preceding and succeeding that frame, resulting in an input feature vector of size $1\times1450$.

\section{Proposed Neural Network Models}
\label{sec:pagestyle}


CNN architectures have been successfully applied on a 2D image arrays \cite{lecun2015deep}. It consists of two typical operations: convolution and pooling. Convolutional layers are responsible for mapping, into their units, detected features from receptive fields in previous layers. This is referred to as a feature map and is the result of a weighted sum of the input features passed through a non-linearity such as ReLU \cite{lecun2015deep}. A pooling layer will typically take the maximum or average of a set of neighboring feature maps, reducing dimensionality by merging semantically similar features. The CNN model proposed here has two convolutional layers with $32$ filters each. In the first layer $25\times30$ filters are used, followed by a $2\times2$ max pooling. A dropout ($0.2$) is used as a regularizer before the next two convolutional layers, which has $64$ $3\times3$ filters each. After the second layer, a $2\times2$ max pooling is applied prior to another dropout ($0.2$). A fully-connected layer of dimensionality $64$ is then used prior to the output unit. We adopted ReLU as an activation function within the hidden units and a learning rate of 0.0001.

As the second architecture, a multilayer perceptron (MLP) is adopted. Such a DNN learns a better feature representation by mapping the input features into a linearly separable feature space \cite{lecun2015deep}. This is achieved by successive linear combinations of the input variables, $z_i=w_ix_i+b_i$, where $w_i$ and $b_i$ are weights and biases, followed by a non-linear activation function. Our first DNN architecture has $400$ input units, followed, respectively, by $200$ and $100$ units in the first and second hidden layers. We used the same activation function, dropout and learning rate adopted for the CNN. The second proposed DNN model receives a feature vector of size $1\times1450$ and has four fully connected layers with $1024$ hidden units each. We adopted, respectively, $0.5$ and $0.0004$ as dropout and learning rate. Adam is used as an optimization algorithm for both architectures. 

Such neural network models require a fixed length of the feature vectors, while the duration of the evaluated audio signal varies. This problem can be addressed either by computing statistics of the features before sending them to the neural network (e.g. i-vectors), or by feeding the neural network with a fixed length of extracted vectors multiple times until the audio file ends, while computing statistics across the timeline. The mean or the mode is typically used, but it is also possible to have an additional classifier, such as the extreme learning machine (ELM)~\cite{HuangZhuSiew2006}, adopted in this work.

\section{Experimental Setup}
\label{sec:Setup}

We compared the performance of our proposed methods to three speech quality metrics. PESQ is adopted as a benchmark as it is one of the most widely used instrumental quality measures. We also included two non-intrusive measures as a benchmark: the speech-to-reverberation energy ratio (SRMR) \cite{falk2010non}, and the ITU-T Recommendation P.563 \cite{ITU_T_P863}. The SRMR has shown to be a good candidate for estimating speech quality and intelligibility, outperforming PESQ and ITU-T P.563, especially in reverberant and dereverberated speech.

The performance of the tested algorithms are compared using two criteria: Pearson's correlation ($\rho$), and mean squared error (MSE). The data is divided in $70$\% for training, $15$\% for validation and hyperparameters optimization, and the other $15$\% for testing. All presented results are based on estimations of the MOS from the test set and the respective MOS attained from subjective scores.

\section{Experimental Results}
\label{sec:results}

\begin{table}[th]
\centering
\scalebox{1}{
\begin{tabular}{c c c c c c c c c c}
 \hline
      Model & $\rho$ & MSE \\
 \hline
 \hline
PESQ & 0.70 & 0.25  \\
SRMR & 0.60 & 0.31 \\
P.563 & 0.55 & 0.36 \\
Constant Q (Spectrum) + CNN & 0.72  & 0.30 \\
i-vector + DNN & 0.78 & 0.22 \\
Mel-Frequency + DNN & 0.86 & 0.18  \\
Mel-Frequency + DNN + ELM & \textbf{0.87} & \textbf{0.15}\\
\hline
\end{tabular}}
\label{tbl:mos_results}
\caption{Results of MOS estimation}
\end{table}

The results are presented in Table 1. The first three lines are the benchmarks, followed by the CNN trained on constant Q  spectral. The DNN using i-vector as a feature set is followed by the results from DNN using Mel-frequency features. The best performance for both evaluation parameters is achieved by the Mel-frequency+DNN+ELM algorithm. Its Pearson's correlation of $0.87$ and MSE of $0.15$ far exceed the non-intrusive standard P.563 with $0.55$ and $0.31$ respectively. PESQ was also surpassed by the DNN+ELM approach. Overall, all the proposed models outperformed the benchmark ones. This is due, in great part, by the fact that the proposed DNN models were able to capture human factors potentially neglected by the standard methods as it can be observed in Fig~\ref{fig:distribution}, where the PESQ distribution seems to be more aligned with the SNR rather than to the human perception, represented by the MOS distribution. 

\section{Conclusion}
\label{sec:conclusion}

In this work, we developed a realistic audio quality dataset based on crowd-sourcing labelling. We also propose three neural network-based approaches for estimating MOS. All models are non-intrusive and their performances are compared to three instrumental measures: PESQ, ITU-T P.563, and SRMR. Results show that all of the proposed approaches outperform the other instrumental measures. The fully connected model using Mel-frequency features as input provided the best correlations and lowest mean squared errors, followed by the DNN model combined with i-vector and the CNN model combined with the constant Q spectrum. As future work, we will evaluate the proposed methods on an extended dataset with network impairments. We will also consider training a DNN model using the raw signal.

\bibliographystyle{IEEEbib}
\bibliography{MyRefs}

\noindent

\end{document}